\documentclass[12pt,leqno]{article}
\usepackage{amsmath,amsfonts}
\usepackage{graphicx}
\usepackage{amsthm,amssymb,bm}
\usepackage{setspace}

\begin{document}
\title{Superintegrable three body systems in one dimension and generalizations}
\author{Claudia Chanu \, Luca Degiovanni \, Giovanni Rastelli \\ \\ Dipartimento di Matematica, Universit\`a di Torino. \\ Torino, via Carlo Alberto 10, Italia.\\ \\ e-mail: claudiamaria.chanu@unito.it \\ luca.degiovanni@gmail.com \\ giorast.giorast@alice.it }
\maketitle

\bigskip

\begin{abstract}
Superintegrable Hamiltonian systems describing the interactions among three point masses on a line have been described in [\ref{E1}]. Here, we show examples of how the approach of above can be  extended  to a higher number of particles on a line and on higher dimensional manifolds. This paper is a slightly extended version of a poster presented at the XVI ICMP held in Prague, 3-8 August 2009. 

\end{abstract}

\paragraph*{A class of superintegrable three-body systems on the line} (a short summary of [\ref{E1}]). Natural Hamiltonian three-body (mass points) systems on a line, with positions $x^i$ and momenta $p_i$ can naturally be interpreted as one point system in the three-dimensional Euclidean space with Cartesian coordinates and momenta $(x^i,p_i)$. The masses of the points can be considered unitary, the case of different positive masses being always reducible to the former [2]. Let $X_i=x^i-x^{i+1}$, $\mathbf{\omega}=(1,1,1)$ and let $(r, \psi,z)$ be cylindrical coordinates of axis parallel to $\mathbf{\omega}$. The Hamiltonians with scalar potentials of the form
\begin{equation}\label{E0}
V=\sum_i {\frac 1{X_i^2}F_i\left(\frac {X_{i+1}}{X_i},\frac {X_{i+2}}{X_i}\right)} =\frac {F(\psi)}{r^2},
\end{equation}
admit the four independent quadratic first integrals
$$
H=\frac 12\left(p_r^2+\frac 1{r^2}p_\psi^2+p_z^2\right)+\frac {F(\psi)}{r^2},\ \ \
H_1=\frac 12p_\psi^2+F(\psi), \ \ \ H_2=\frac 12 p_z^2,
$$
$$
H_3=\frac 12\left(rp_z-zp_r\right)^2+\frac {z^2}{r^2}H_1.
$$
The first three first integrals allow integration of the Hamilton-Jacobi equation by separation of variables and determine the radial motion common to any choice of $F$.
Examples are Calogero and Wolfes potentials
$$
V_C=\sum_{i=1}^3\frac {k_i}{X_i^2}=\frac k{\left( r\, \sin 3\psi \right)^2},\ \ 
V_W=\sum_{i=1}^3\frac {h_i}{\left(X_i-X_{i+1}\right)^2}= \frac h{\left( r\,\cos 3\psi \right)^2},
$$
with $(h,h_i,k,k_i \in R)$. Remarkably, a rotation of angle $\alpha=\frac \pi 6$ around the axis $\Omega$ of the cylindrical coordinates, i.e. a phase shift $\psi\longrightarrow \psi+\alpha$, makes the separated equations for $V_C$, a two-body interaction,  and $V_W$, a three-body interaction, to coincide. Therefore, by reading the potentials as functions of $X_i$, the corresponding interactions among points on the line can be considered as equivalent. Potentials as $V_C$ and $V_W$ admit a further independent cubic first integral making them maximally superintegrable. In general, any potential of the form $V=\frac k{[r\, \sin (2n+1)\psi]^2}$
seems to admit the fifth first integral 
$$
\sum_{\sigma =0}^{n}\sum_{i=0}^{2\sigma+1} {\frac{A_\sigma ^i }{r^{2n+1-i}}\left(\frac{2k}{\sin ^2 (2n+1)\psi}\right)^{n-\sigma}\frac {d^l\left(\cos (2n+1)\psi\right)}{d\psi^l} p_r^ip_\psi^l}
$$
with $l=2\sigma+1-i$ and
$$
A_\sigma^i=\frac{(-1)^{2n-\sigma}}{(2n+1)^{2\sigma+1-i}} \; \left( \begin{matrix} {2n+1} \cr i \end{matrix}\right) \left( \begin{matrix} [(2n+1-i)/2] \cr [(2\sigma+1 -i)/2]\end{matrix} \right),  
$$
where $\left(^a _b\right) =\frac {a!}{b!(a-b)!}$ denotes the Newton binomial symbol and $[a]$ the greatest integer  $\leq a$.

\paragraph*{$n$ points on a line}

Any  homogeneous function of degree -2 in $X_i=x^i-x^{i+1}$, $i=1..n-1$, can be written in the form
\begin{equation}\label{E1}
V=\frac 1{X_1^2}F\left(\frac {X_2}{X_1},\ldots,\frac{X_{n-1}}{X_1}\right)=\frac 1{r_{n-1}^2}\Phi(\psi_1,\ldots, \psi_{n-2}),
\end{equation}
and viceversa, where $\tan \psi_i=\frac {X_{i+1}}{X_1}$ and $(r_{n-1},\psi_1,\ldots,\psi_{n-2},u)$ are spherical-cylindrical coordinates  in $E^n$.

\proof Let $(z^j)$ coordinates in $R_n$ defined by
\begin{eqnarray*}
z^j=\frac 1{\sqrt{j(j+1)}}(x^1+x^2+\ldots +x^j-jx^{j+1}) \quad j=1\ldots n-1\\
z^n=\frac 1{\sqrt{n}}(x^1+\ldots+x^n),
\end{eqnarray*}
they are Cartesian coordinates equioriented with $(x^i)$. It follows
$$
\frac {z^j}{z^1}=\frac{\sqrt{2}}{\sqrt{j(j+1)}}(1+2\frac {x^2-x^3}{x^1-x^2}+\ldots+j\frac{x^j-x^{j+1}}{x^1-x^2}).
$$
Therefore
$$
\frac 1{X_1^2}F\left(\frac {X_2}{X_1},\ldots,\frac{X_{n-1}}{X_1}\right)=\frac 1{(z^1)^2}\Phi(\frac{z^2}{z^1},\ldots, \frac{z^{n-1}}{z^1}).
$$
Moreover,
$$
(z^1)^2+\ldots+(z^{n-1})^2=(z^1)^2\left( 1 + \left(\frac{z^2}{z^1}\right)^2+\ldots +\left(\frac{z^{n-1}}{z^1}\right)^2\right). 
$$
It is evident that any homogeneous function of degree -2 in $X_i$ can be written as the central term of (\ref{E1}) and that both expressions in (\ref{E1}) are homogeneous  of degree -2 in $X_i$. Zeros of $(X_i)$ are obviously not considered here, corresponding to collision configurations of the particles.

\endproof

 Natural systems on $E^n$ with potential of the form (\ref{E1}) can be interpreted as natural  systems of $n$ points on a line. Phase shifts $\psi_i\longrightarrow\psi_i+\alpha_i$, $0\leq \alpha_i< \pi$, determine equivalence classes of interactions on the line as seen in dimension three.

\paragraph*{Evans systems embedded in dimension 4}

All superintegrable potentials with at least four quadratic first integrals in $E^3$ are listed in [1]; those in the form (\ref{E1}) are, up to isometries,
$$
V_1=\frac {F(\psi_1)}{(\rho\sin \psi_2)^2}, \ V_2=\frac k{(\rho \cos \psi_2)^2}+\frac {F(\psi_1)}{(\rho \sin \psi_2)^2},\ V_3=\frac {k\cos \psi_2+F(\psi_1)}{(\rho \sin \psi_2)^2},
$$
with  $\rho^2=x^2+y^2+z^2$. Some other particular cases are possible, but no undetermined functions are then included; for example
$$
V_4=\frac k{\rho^2}\left[k+\frac {k_1}{(\cos\psi_2)^2}+\frac 1{(\sin\psi_2)^2}\left(\frac {k_2}{(\cos \psi_1)^2}+\frac {k_3}{(\sin \psi_1)^2}\right)\right].
$$
All the potentials of above, trivially extended in dimension four by adding $\frac 12p_u^2$ to the three-dimensional Hamiltonian $H$, admit six independent quadratic first integrals in $E^4$: the new Hamiltonian $\frac 12p_u^2+H$, 
$$
H_1=\frac 12\left(p_{\psi_2}^2+\frac 1{(\sin \psi_2)^2}p_{\psi_1}^2\right)+\rho^2 V_i,
$$
the three other "old" first integrals of $H$ and 
$$
H_5=p_u^2,\ \ H_6=\frac 12(up_\rho-\rho p_u) ^2+\frac {u^2}{\rho^2}H_1.
$$
It is always possible to write the functions of above in term of distances between four points on a line; again, phase shifts in $\psi_i$ determine equivalence classes of interactions on the line. Then, for $n=4$, it is determined a class of superintegrable four-body systems on the line with $2n-2$ quadratic first integrals.

\paragraph*{Many-body systems in higher-dimensional manifolds}

The approach of above can be easily generalized to systems of $n$ points in $m$-dimensional Euclidean spaces. Let $(q^i_r,p_i^r)$ denote the $i$-th canonical euclidean coordinates of the $r$-th of $n$ points in a $m$ dimensional Euclidean space. If we require that the total momentum of the system is a constant of the motion, then, the Hamiltonian depend on  the differences of the configuration coordinates of the points only: $H=H\left(p_r,(q^i_r-q^i_s)\right)$.

The previous Hamiltonian system can be interpreted as a single point Hamiltonian in a $m\cdot n$ dimensional Euclidean space by introducing canonical cartesian coordinates $(y_i,x^i)$ by
$$
y_{m(r-1)+i}=p^r_i, \qquad x^{m(r-1)+i}=q_r^i, \qquad r=1,\dots, n, \quad i=1, \dots, m.
$$
Hence, the $m\cdot n$ dimensional Hamiltonian is $H=\frac 12 \Sigma_{j=1}^{m\cdot n}y_j^2+V(x^j)$
The constance of the total momentum is equivalent to the existence of the $m$ first integrals
$\Sigma_{r=1}^np^r_i$, $(i=1,\dots, m),$
that, in $(y,x)$, is equivalent to the invariance of $H$ with respect to the $m$ vectors $\omega_i$ of components 
$$
\omega_i^j=\delta ^j_{r\cdot m+i}, \qquad r=1, \dots, n,
$$
where $\delta$ is the Kroenecker symbol. For example, $m=2$, $n=3$ corresponds to three points in an Euclidean plane $M$ and in cartesian coordinates $(y_i,x^i)$, $i=1,\dots, 6$, we have
$$
\omega_1=(1,0,1,0,1,0),\ \ \ \omega_2=(0,1,0,1,0,1).
$$ 
By imposing invariance with respect to the two vectors
$$
\omega_3=(1,0,-1,0,0,0),\ \ \ \omega_4=(0,1,0,-1,0,0),
$$ 
a two-dimensional plane in $R^6$ is determined. Here, a potential of the form (\ref{E0}) can be introduced. The natural Hamiltonian is
$$
H=\frac 12(p_1^2+p_2^2+p_3^2+p_4^2+p_r^2)+\frac 1{r^2}H_1,
$$
in Cylindrical coordinates $(r,\psi,u_i)$ with "axis" generated by the $(\omega_i)$, and its nine independent quadratic first integrals are: $H_1=\frac 12p_\psi^2+F(\psi)$,
$$
H_i=p_i^2, \ \ H'_i=\frac 12(rp_i-u_ip_r)^2+\frac{u_i^2}{r^2}H_1,\ \ i=1,\ldots, 4.
$$
The potential $V$ in the original plane $M$ assumes the form
$$
V=\frac 1{X_1^2}F_1\left(\frac{X_2}{X_1}\right)+\frac 1{X_2^2}F_2\left(\frac{X_1}{X_2}\right), 
$$
where  $X_1=x^1+x^3-2x^5$,  $X_2=x^2+x^4-2x^6$.

\paragraph*{Conclusion and further readings}

In this paper we show just some partial results of a study in progress. Detailed references can be found in \cite{E1}. Recent papers about the arguments considered here, or very cose to,  are    \cite{E2} and \cite{E3}.

\end{document}